\documentstyle[a4wide,epsf,11pt,titlepage]{article}

\newcounter{nref}
\newcommand{\bbib}{%
  \renewcommand{\refname}{\large\bf References}%
  \begin{thebibliography}{99}%
  \setcounter{enumiv}{\thenref}}
\newcommand{\ebib}{%
  \end{thebibliography}%
  \setcounter{nref}{\arabic{enumiv}}}
\newcommand{\head}[3]{%
  \setcounter{nref}{0}%
  \thispagestyle{empty}%
  \section*{\LARGE\bf #1}%
  \stepcounter{section}%
  \addcontentsline{toc}{section}{#1}%
  \large%
  #2\\\vspace{0.1pt}\\%
  #3%
  \normalsize%
  \bigskip}

\def\Msol{{\rm M_{{\odot}}}}

\newcommand{\Ox}{\rm{^{16}O}}

\newcommand{\Si}{\rm{^{28}Si}}

\newcommand{\Ni}{\rm{^{56}Ni}}

\begin{document}


\head{The First Hours of a Core Collapse Supernova}
     {K.\ Kifonidis$^1$, T.\ Plewa$^{2,1}$, H.-Th.\ Janka$^1$, E.\ M\"uller$^1$}
     {$^1$ Max-Planck-Institut f\"ur Astrophysik,
      Karl-Schwarzschild-Strasse 1, D-85740 Garching, Germany\\ $^2$
      Nicolaus Copernicus Astronomical Center, Bartycka 18, 00716
      Warsaw, Poland}

\subsection*{Abstract}

New two-dimensional, high-resolution calculations of a core collapse
supernova in a $15\,\Msol$ blue supergiant are presented, which cover
the entire evolution from shock revival until the first few hours of
the explosion.  Explosive nucleosynthesis, its dependence upon
convective mixing during the first second of the evolution and the
growth of Rayleigh-Taylor instabilities at the composition interfaces
of the progenitor star are all modeled consistently and allow for a
comparison with observational data. We confirm our earlier findings,
that the perturbations induced by neutrino driven convection are
sufficiently strong to seed large-scale Rayleigh-Taylor mixing and to
destroy the onion-shell structure of the stellar He-core. As in our
earlier calculations, the strong deceleration of the nickel clumps in
the layers adjacent to the He/H interface suggests that the high
velocities of iron-group elements observed in SN 1987\,A cannot be
explained on the basis of currently favored progenitor models.
Possible solutions to this dilemma and the implications of the mixing
for type Ib explosions are briefly discussed.

\subsection{Introduction}

Recent observations of the Cas~A remnant by NASA's Chandra X-ray
observatory \cite{Kifonidis.Hughes+00} appear to directly disclose for
the first time the nature of the violent processes which are
responsible for the synthesis of heavy and intermediate mass elements
in core collapse supernovae. The spatial separation of silicon and
iron emission observed by Hughes et al. \cite{Kifonidis.Hughes+00} has
been interpreted as the result of a large-scale overturn during the
earliest phases of the explosion of the Cas~A progenitor.  
Together with the numerous indications of
strong mixing in e.g. SN 1987\,A \cite{Kifonidis.Arnett} and SN
1993\,J \cite{Kifonidis.Spyromilio} these new observations begin to
assemble into a picture in which hydrodynamic instabilities in
supernova explosions are inevitable and must be regarded as a key to
an understanding of these events. Thus multidimensional hydrodynamic
simulations are urgently required which must cover the entire
evolution from shock formation until shock breakout through the
stellar photosphere.

In \cite{Kifonidis.LETTER} we have
reported on first preliminary results of such calculations and found
that neutrino driven convection is able to seed strong Rayleigh-Taylor
mixing at the Si/O and (C+O)/He interfaces of the SN 1987\,A
progenitor model of \cite{Kifonidis.WPE88} within only about a minute
after core bounce. The stellar metal core was found to have been
completely shredded only five minutes after bounce and high-velocity
clumps of newly synthesized elements were observed to be ejected up 
to the outer edge of the helium core. However, these were substantially
decelerated in a dense shell that formed at the He/H interface after
the supernova shock entered the hydrogen envelope of the star.
In this contribution we present more refined calculations for the same
presupernova star, which cover a longer period of the evolution 
and which are used to test the numerical sensitivity of our
earlier results.

\subsection{Models}

In our new calculations we have accomplished to overcome the numerical
problems due to ``odd-even decoupling'' that were observed in
\cite{Kifonidis.LETTER}, to increase the spatial resolution and to
include gravity into the adaptive mesh refinement (AMR) calculations
of the Rayleigh-Taylor growth phase, which now cover the time span
from 820\,ms up to more than 20\,000\,s after bounce.  Different from
\cite{Kifonidis.LETTER} we have also used a new, somewhat more
energetic explosion model ($E_{\rm expl} = 1.8\times 10^{51}$\,ergs as
compared to $E_{\rm expl} = 1.5\times 10^{51}$\,ergs).  Note, however,
that since gravity was neglected in our old AMR calculations, the old
model resulted in a kinetic energy at infinity of $E_{\rm expl}
\approx 1.8\times 10^{51}$\,ergs which is by about 20\% {\em larger\/}
than in the new simulations. This, of course, resulted in a slower
overall expansion in our new calculations, giving the Rayleigh-Taylor
instability somewhat more time to grow before the clumps reached the
outer edge of the He core.

Not being affected by noise due to odd-even decoupling, our new
simulations show qualitative differences in the growth of the
Rayleigh-Taylor instability. Seeded by neutrino driven convection in
the deeper layers of complete silicon burning, the instability starts
to grow about 50\,s after bounce at the Si/O interface on much smaller
angular scales (of the order of $\leq 1^{\circ}$) than found in
\cite{Kifonidis.LETTER}.  Superposed upon the resulting mushrooms,
which grow out of the highest frequency perturbations that can still
be resolved on our grid, are long-wavelength perturbations of the
entire interface which are caused by the convective blobs beneath it.
From these perturbations evolve initially cusps and subsequently
fully-grown fingers, which start to perturb also the unstable O/He
interface.  Fig.~\ref{kifonidis.fig1} shows the situation 1170\,s
after core bounce when the instabilities at both interfaces are
already fully developed.  Depicted are the spatial distribution of the
mass density as well as the partial densities of oxygen, silicon and
nickel.  Fine-grained filaments of dense material which includes
silicon as well as newly synthesized $\Ni$ (and other nuclides) are
visible, which are embedded in broader oxygen fingers that penetrate
through the He-core. However, no overturn of the $\Si$ and $\Ni$ rich
layers as suggested by \cite{Kifonidis.Hughes+00} is found in which
the $\Ni$ is mixed much farther out than the silicon. Both elements
are rather evenly distributed throughout the He-core.

We find maximum $\Ni$ velocities around 3000\,km/s before the clumps
penetrate into the dense shell at the He/H interface (see
\cite{Kifonidis.LETTER}) 1600\,s after bounce, and their velocities
decrease to about 1000\,km/s at the end of our calculations. Both of
the quoted values are substantially lower than in our earlier
simulations.  The dense shell is caused by the compression of the
post-shock matter behind the decelerating shock, which encounters a
flatter slope of the density profile once it crosses the He/H
interface and enters the hydrogen envelope.  The strong flattening of
the density gradient is a generic feature in all presupernova models
proposed for SN 1987\,A and makes dense shell formation inevitable
during the explosion.  Thus our new calculations underline that on the
basis of presently favored progenitor and current multidimensional
explosion models, it is not possible to explain the high nickel
velocities in two-dimensional calculations.

\begin{figure}[ht]
  \caption{Spatial distribution of the density (left) and the partial
  densities (right) of $\Ox$ (blue), $\Si$ (green, turquoise, white),
  and $\Ni$ (red, pink) in the inner He-core of the exploding star
  1700 seconds after bounce. At this time, the supernova shock is
  already propagating through the hydrogen envelope. The onion shell
  structure of the core has been completely shredded by
  Rayleigh-Taylor instabilities at the Si/O and (C+O)/He
  interfaces. $\Si$, $\Ni$ and all other products of explosive oxygen
  and silicon burning are localized in dense, rapidly expanding clumps
  which are propelled through the He core.}
\label{kifonidis.fig1}
\end{figure}

\subsection{Conclusions}

Though in detail differences as compared to \cite{Kifonidis.LETTER}
are found, the main results and conclusions of our earlier work remain
unchanged. We confirm that neutrino driven convection succeeds in
seeding large-scale mixing processes in the exploding star which
destroy its onion-shell structure on  a time-scale of {\em
minutes\/}.  Although our calculations do not yield the kind of
overturn of iron and silicon-rich layers claimed to be present in
Cas~A by \cite{Kifonidis.Hughes+00}, they appear to yield a natural
explanation for the mixing occuring in type~Ib supernovae (see also 
\cite{Kifonidis.Woosley_Eastman97}). 

The situation is much more unclear in case of SN 1987\,A.  Definitely,
a reliable modeling of the deceleration of the nickel clumps in the
dense shell at the He/H interface has to take into account the
different drag that genuinely three-dimensional ``mushrooms''
experience as compared to two-dimensional tori
\cite{Kifonidis.Kane+00}. Thus three-dimensional calculations are
required before one will be able to draw definite
conclusions. However, the present calculations indicate that standard
stellar evolution models for the progenitor of this supernova might
have to be abandoned in favor of merger models, which appear to be the
most likely explanation for such large differences of the
hydrogen-envelope structure as our hydrodynamic simulations require.

An alternative explanation might, however, be sought in ``missing
physics'' in the modeling of the explosion itself. Large-scale
asymmetries, e.g. produced by jets \cite{Kifonidis.Fryer_Heger},
\cite{Kifonidis.Nagataki}, may accelerate the products of complete
silicon burning to velocities that might be sufficient to propel these
elements up to the stellar hydrogen envelope.

\subsection*{Acknowledgements}

It is a pleasure to thank S. E. Woosley for providing us with his
progenitor models and for pointing out the importance of the mixing
for the spectra of Type~Ib explosions.


\end{document}